\documentclass[twocolumn,prb,showpacs,multicol,amsmath,amssymb]{revtex4}
\usepackage[dvips]{graphicx}
\usepackage{graphicx}
\usepackage{dcolumn}
\usepackage{bm}
\usepackage{graphics}
\usepackage{epsfig,color}
\usepackage[normalem]{ulem} 
\usepackage{soul}

\newcommand{\be}{\begin{equation}}
\newcommand{\ee}{\end{equation}}
\newcommand{\bea}{\begin{eqnarray}}
\newcommand{\eea}{\end{eqnarray}}

\newcommand{\bS}{\bf S}

\begin{document}
\title{Markovian and Non-Markovian dynamics in the one-dimensional transverse-field XY model}
\author{Z. Saghafi$^{1}$}
\email[]{Z.Saghafiv@gmail.com}
\author{S. Mahdavifar$^{1}$}
\author{E. Hosseini Lapasar$^{1}$,$^{2}$}
\affiliation{$^{1}$Department of Physics, University of Guilan, 41335-1914 Rasht, Iran}
\affiliation{$^{2}$Department of Chemistry and Molecular Material Science, Graduate School of Science, Osaka City University, Osaka 558-8585, Japan}
\begin{abstract}
We consider an anisotropic spin-1/2 XY Heisenberg chain in the presence of a transverse magnetic field. Selecting the nearest neighbor pair spins as an open quantum system, the rest of the chain plays the role of the structured environment. In fact, the aforementioned system is used as a quantum probe signifying nontrivial features of the environment with which is interacting. We use a general measure that is based on the trace distance for the degree of non-Markovian behavior in open quantum systems. The witness of non-Markovianity takes on nonzero values whenever there is a flow of information from the environment back to the open system. We have shown that the dynamics of the system with isotropic Heisenberg interaction is Markovian. A dynamical transition into the non-Markovian regime is observed as soon as the anisotropy, $\gamma$, is applied. At the Ising value of the anisotropy $\gamma=1.0$, all the information flows back from the environment to the system. The additional dynamical transition from the non-Markovian to the Markovian is obtained by applying the transverse magnetic field. In addition, we have focused on the time evolution of the Loschmidt-echo return rate function. It is found that a non-analyticity can be seen in the time evolution of the Loschmidt-echo return rate function exactly at the critical points where a dynamical transition from the Markovian to the non-Markovian occurs.
\end{abstract}
\maketitle

\section{Introduction}\label{sec1}
The concept of non-Markovian dynamics is known as the main subject in the theory of open quantum systems\cite{Breuer07-1, Rivas14}. In reality, no quantum system can be completely isolated from its environment. Thus, a complete description of a system requires to include the effect of the environment. Often, an open quantum system is a subsystem of some larger system composed of open system and its environment.
Non-Markovian processes feature a flow of information from the environment back to the open system, which implies the presence of memory effects and represents the key property of the non-Markovian quantum behavior. The unidirectional flow of the information from the open system to the environment is known as Markovian which was significantly successful in the frontier of quantum optics. In the Markovian process, an open system irretrievably loses information to its environment. 
Many-body condensed matter systems are the best hotbed for the non-Markovian process, describing strong interaction between system and environment. In this topic, researches have considered different interactions between particles and environments. Especially, the studies on the dynamics of open quantum systems have recently been focused on the non-Markovian environments\cite{Santos06, Bellomo07, Piilo08, Appolaro11, Madsen11, Liu11, Franco12,  Huelga12, Barnes12, Haikka12, Laine12, Franco13, Xu13, Lorenzo13, Addis14, Orieux15}. The concurrence decays exponentially and asymptotically in the Markovian dynamics\cite{Santos06}. A procedure obtaining the dynamics of a system of $N$ independent bodies, each locally interacting with an environment is shown\cite{Bellomo07}. The quantum jump method is presented for obtaining the dynamics of the open quantum systems that interact with the non-Markovian environment \cite{Piilo08}. Revivals of entanglement are the most noticeable result observed for open systems in contact with non-Markovian environments\cite{Bellomo07, Franco12, Franco13, Xu13}.  It is also shown that enlarging an open system could change the dynamics from the Markovian to the non-Markovian\cite{Laine12}. Also, it is possible to induce a full dynamical transition from the Markovian to the non-Markovian for the two-level system by controlling parameters such as the mismatch between the energy of the two-level system and the spin environment\cite{Lorenzo13}. A comparison has been done between several recently proposed non-Markovianity measures for the single and the composite open quantum systems\cite{Addis14}.  
Most studies are related to the dynamics of a single spin-1/2 particle as an open quantum system.  The dynamics of the Loschmidt echo is shown to be directly linked to the information flux between a spin-1/2 particle and its environment\cite{Haikka12}. On the other hand, the dynamics of a qubit coupled to a spin chain environment is studied \cite{Appolaro11}. The spin chain environment is described by a spin-1/2 XY model in a transverse magnetic field. There is a specific point in the parameter space of the system, where the qubit dynamics is Markovian. Two regions which are separated by this point triggers two totally different dynamical behaviors. In other work, A system of  $N$-particle is separated into two parts; a single qubit and an environment which strongly coupled to the qubit\cite{Marko11}. It is found that the contribution due to energy density is responsible for the non-Markovian effects even in the limit of an infinite environment. Recently, the dynamics of an open quantum system containing a pair of nearest neighbor spins coupled to a spin-1/2 XX chain environment is studied\cite{Mahmoudi17}. It is found that the dynamical transition from the Markovian to the non-Markovian regime occurs by increasing the three-spin interaction.
\begin{figure}[t]
\centerline{\psfig{file=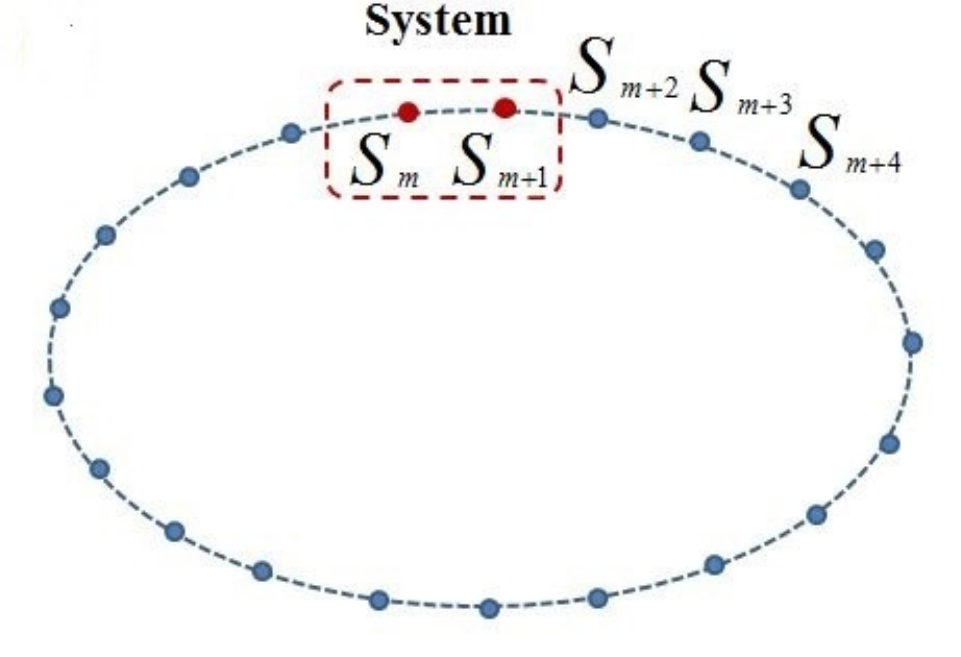,width=2.7in}}
\caption{(color online). The schematic diagram of the open quantum system and its environment\cite{Mahmoudi17}.}
\label{Fig-0}
\end{figure}
From recent studies, a general theory is made which rigorously define the border between the Markovian and the non-Markovian quantum dynamics\cite{Breuer09, Breuer12}.  Mathematical characterizations of the non-Markovian behavior are provided by approaches in this theory. A very efficient tool as a kind of witness for the non-Markovianity is typically distinguishability between two different initial quantum states, namely $\rho_{1}$ and $\rho_{2}$. This is a measure of the distance of two quantum states known as the trace distance\cite{Breuer09}, which quantifies the distinguishability of two different initial quantum states. In principle, the gain or loss of quantum information can be quantified through the dynamics of the trace distance between a pair of quantum states of the open system. The trace distance is defined as
\begin{eqnarray}
D(\rho_1, \rho_2)=\frac{1}{2}tr\vert \rho_1-\rho_2 \vert,
\end{eqnarray}
where $\vert A \vert=\sqrt{A^{\dagger}A}$. There is a fact that all completely positive and trace-preserving maps $\lambda$ are contractions for the trace distance,
\begin{eqnarray}
D(\lambda \rho_1, \lambda \rho_2)\leq D(\rho_1, \rho_2),
\end{eqnarray}
which means that the trace distance will not be increased in any completely positive and trace-preserving map. Therefore, when the environment information is returned to the system, the trace distance increases at some times which is known as a clear sign of the non-Markovianity. One should note that the degree of memory effect can be obtained by
\begin{eqnarray}
{\cal N}=max~_{\rho_{1,2}(0)} \int_{\sigma>0}dt \sigma(t,\rho_{1,2}(0)),
\end{eqnarray}
where
\begin{eqnarray}
\sigma(t,\rho_{1,2}(0))=\frac{d}{dt} D(\rho_1(t), \rho_2(t)),
\end{eqnarray}
is the rate of change of the trace distance at time $t$. It should be noted that the time integral is extended over all intervals where the trace distance rises by increasing time. On the other hand, the maximum is taken over all pairs of initial states, which is called an optimal state pair. The optimal state pair leads to the maximal possible backflow of the information during the time evolution of states. Thus, by construction, we have ${\cal N}\neq 0$ if and only if the process is the non-Markovian.

In this paper, we consider an anisotropic spin-1/2 XY  Heisenberg chain in the presence of a transverse magnetic field. Using the Jordan-Wigner transformations, the system is mapped onto the noninteracting fermion system. Applying the Bogoliubov transformations, the Fermionic Hamiltonian is diagonalized. The nearest neighbor pair of spins is selected as an open quantum system, so the others act as an environment (Fig.~\ref{Fig-0}). We show that the dynamics of the system is Markovian in the absence of the transverse magnetic field and the anisotropy.  As soon as the interaction becomes anisotropic, a dynamical transition occurs to the non-Markovian regime. Especially, we show that at the certain Ising value of the anisotropy $\gamma=1.0$, all the information is returned from the environment to the open quantum system. By applying a transverse magnetic field, a new dynamical transition occurs from the non-Markovian to the Markovian regime in the region of the anisotropy $\gamma \leq 1.0$. Due to the strong quantum fluctuations, no dynamical transition is observed in the region of the anisotropy $\gamma>1.0$.  In addition, we focus on the time behavior of the Loschmidt-echo return rate function and explicitly show that the dynamical transition from the Markovian to the non-Markovian is revealed itself as a non-analyticity in the real-time evolution of the Loschmidt-echo return rate function.
The paper is organized as follows. In the forthcoming section, we introduce the model and present our exact analytical results for the density matrix of quantum states. In Sec. III, we argue about the witness of the non-Markovianity and present our analytical results. The results are concluded and summarized in Sec. IV.

\section{THE MODEL}\label{sec2}
The Hamiltonian of an anisotropic spin-1/2 XY chain in the presence of a transverse magnetic field is written as
\begin{eqnarray}
{\cal H}&=&-J \sum_{n=1}^{N}[(1+\gamma){S}_{n}^x {S}_{n+1}^x+(1-\gamma){S}_{n}^y {S}_{n+1}^y]  \nonumber\\
  &-&h\sum_{n=1}^{N}{S}_{n}^{z}~,  
\label{Hamiltonian s}
\end{eqnarray}
where $S_{n}$ is the spin-1/2 operator on the $n$-th site and sum over $n$ that goes from $1$ to $N$, satisfying the periodic boundary conditions. The parameter $J$ denotes the exchange coupling constant, $\gamma\neq0$ is the anisotropy parameter and $h$ is the applied external magnetic field. At the special values of anisotropy parameter $\gamma=0$ and $\gamma=1$, the model reduces to the spin-1/2 isotropic XX and Ising model respectively.  Using the transformation $S_n^{x} \to - S_n^{y}$, $S_n^{y} \to S_n^{x}$, $S_n^z \to S_n^z$, the futures of the system can be reconstructed for $\gamma<0$. This model is exactly solvable. By using the Jordan-Wigner transformation
\begin{eqnarray}
{S}_{n}^{+}&=&a_{n}^{\dagger} \exp(i\pi\sum_{l<n}a_{l}^{\dagger}a_{l})~,  \nonumber\\
{S}_{n}^{-}&=&a_{n} \exp(-i\pi\sum_{l<n}a_{l}^{\dagger}a_{l})~, \nonumber\\
{S}_{n}^{z}&=&a_{n}^{\dagger}a_{n}-\frac{1}{2}~,
\end{eqnarray}
the Hamiltonian is mapped onto a noninteracting spinless fermion model as
\begin{eqnarray}
{\cal H}_{f}&=& \frac{-J}{2} \sum_{n}(a^{\dag}_{n}a_{n+1}+a^{\dag}_{n+1}a_{n}+
\gamma (a^{\dag}_{n}a^{\dag}_{n+1}-a_{n}a_{n+1}))\nonumber \\
&-&h\sum_{n} a^{\dag}_{n}a_{n}~.
\end{eqnarray}
By performing a Fourier transformation into the momentum space as $a_{n}^{\dag} = \frac{1}{\sqrt{N}} \sum _{k=1}  e^{ikn} a_{k}^{\dag}$,
and also employing Bogoliubov transformation, $\beta_{k}^{\dag}=u_k a_k^{\dag}-i v_k a_{-k}$, the diagonalized Hamiltonian is obtained as
\begin{eqnarray}
{\cal H}_{f}=\sum_{k}\varepsilon(k) (\beta_{k}^{\dagger} \beta_{k}-\frac{1}{2}),
\label{Hamiltonian d}
\end{eqnarray}
where $\varepsilon(k)$ denotes the dispersion relation and is related to the Hamiltonian's parameters as 
\begin{eqnarray}
\varepsilon(k) &=&  \sqrt{a(k)^2+4b(k)^2}, \nonumber\\
a(k)&=&-J \cos(k)-h, \nonumber\\
b(k)&=& \frac{J \gamma}{2} \sin(k).
\end{eqnarray}
The Bogoliubov's coefficients are related to the Hamiltonian's parameters as $u_k=\sqrt{\frac{1}{2}+\frac{a(k)}{2 \varepsilon(k)}}$ and  $v_k=\sqrt{\frac{1}{2}-\frac{a(k)}{2 \varepsilon(k)}}$. Now, let us select the nearest neighbor pair spins located at sites $m$ and $m+1$ in the chain system as an open quantum system. It is clear that the rest of the chain plays as its environment. The general form of the density matrix of the mentioned open quantum system in the standard basis is expressed as\cite{Mahmoudi17}
\begin{eqnarray}
\rho= \left(
             \begin{array}{cccc}
               <P^{\uparrow}P^{\uparrow}> & <P^{\uparrow}{\bS}^{-}> & <{\bS}^{-}P^{\uparrow}> & <{\bS}{\bS}^{-}> \\
               <P^{\uparrow}{\bS}^{+}> & <P^{\uparrow}P^{\downarrow}> & <{\bS}^{-}{\bS}^{+}> & <{\bS}^{-}P^{\downarrow}> \\
               <{\bS}^{+}P^{\uparrow}> & <{\bS}^{+}{\bS}^{-}> & <P^{\downarrow}P^{\uparrow}> & <P^{\downarrow}{\bS}^{-}> \\
               <{\bS}^{+}{\bS}^{+}> & <{\bS}^{+}P^{\downarrow}> & <P^{\downarrow}{\bS}^{+}> & <P^{\downarrow}P^{\downarrow}> \\
             \end{array}\nonumber
           \right). \label{density matrix1}
\end{eqnarray}
where $P^{\uparrow}=\frac{1}{2}+S^{z}, P^{\downarrow}=\frac{1}{2}-S^{z}$. The brackets denote the expectation  values at time $t$ and  $S^{\pm}= S^{x}\pm i S^{y}$. By using the Jordan-Wigner transformation, the reduced density matrix for the open quantum system will be given by
\[
\rho =
\left( {\begin{array}{cccc}
  X^{+}  & 0 &  0 & 0\\
    0 &  Y^{+}  & Z^{*} & 0\\
 0 & Z & Y^{-}  & 0\\
 0 & 0 & 0 & X^{-}
 \end{array} } \right),
\]
where
\begin{eqnarray}
X^{+}&=&\langle n_{m}(t) n_{m+1}(t)\rangle, \nonumber\\
X^{-}&=&\langle 1-n_m(t)- n_{m+1}(t)+n_m(t) n_{m+1}(t)\rangle, \nonumber\\
Y^{+}&=&\langle n_m(t)(1-n_{m+1}(t))\rangle, \nonumber\\
Y^{-}&=&\langle n_{m+1}(t)(1-n_{m}(t))\rangle, \nonumber\\
Z^{*}&=&\langle a_{m}^{\dagger}(t) a_{m+1}(t) \rangle,
\end{eqnarray}
where $n_m=a_{m}^{\dagger}(t)a_{m}(t)$.\\
Using the solution of the retarded Green’s function\cite{Fetter71}, $X^{+}$ approximately is obtained as 
\begin{eqnarray}
X^{+}&=&\langle a^{\dag}_{m} (t) a_{m} (t)\rangle~\langle a^{\dag}_{m+1}(t) a_{m+1}(t)\rangle \nonumber \\
&-&\langle a^{\dag}_{m} (t) a_{m+1}(t)\rangle~ \langle  a^{\dag}_{m+1}(t) a_{m}(t)\rangle\nonumber \\
&=&\langle n_m \rangle \langle n_{m+1} \rangle - Z^{*} Z.
\end{eqnarray}
One should note that the mentioned approximation preserves the positivity and unit trace condition of $\rho$. For example let us consider the $X^{+}=\langle n_m \rangle \langle n_{m+1} \rangle - Z^{*} Z$. It is clear that $0\leq \langle n_m \rangle \leq 1$ and $-0.25\leq Z^{*} Z \leq 0.25$. Even For the vacuum state, the number operator vanishes, but in this case $Z^{*} Z$ will be also zero and then the positivity is preserved. {(As evidence of this claim, the plot of $X^{+}$ as a function of time is placed in the appendix)}. The time evolution of the original spin operators or equivalently the spinless fermions are the key to determine the time evolution of the quantum state of the system. Applying, the time evolution operator, $U(t)=e^{\frac{-it}{\hbar}\sum_{k}\varepsilon(k) \beta_{k}^{\dagger}\beta_{k}}$, the time-dependent fermion creation operator is obtained as
\begin{eqnarray}
a^{\dag}_{k}(t)=u_k e^{i t \varepsilon(k)} \beta_{k}^{\dagger}+i v_k e^{-i t \varepsilon(-k)} \beta_{-k}.
\label{rel}
\end{eqnarray}
It should be noted that since the dispersion relation is an even function of the momentum, therefore, $\varepsilon(k)=\varepsilon(-k)$.

\section{Trace distance}\label{sec3}
In this section, we will study the evolution of the distinguishability between the pair of reduced states which describing the information flows between the open quantum system and its environment. It is known that the trace distance is bounded as $0\leq D(\rho_1, \rho_2)\leq1$, where $D(\rho_1, \rho_2)=0$ if and only if
the pair states are the same, and  $D(\rho_1, \rho_2)=1$, if and only if the pair states are orthogonal. One should note that a trace preserving quantum
operation can never increase the distinguishability of any pair states.
\begin{figure}[t]
\centerline{\psfig{file=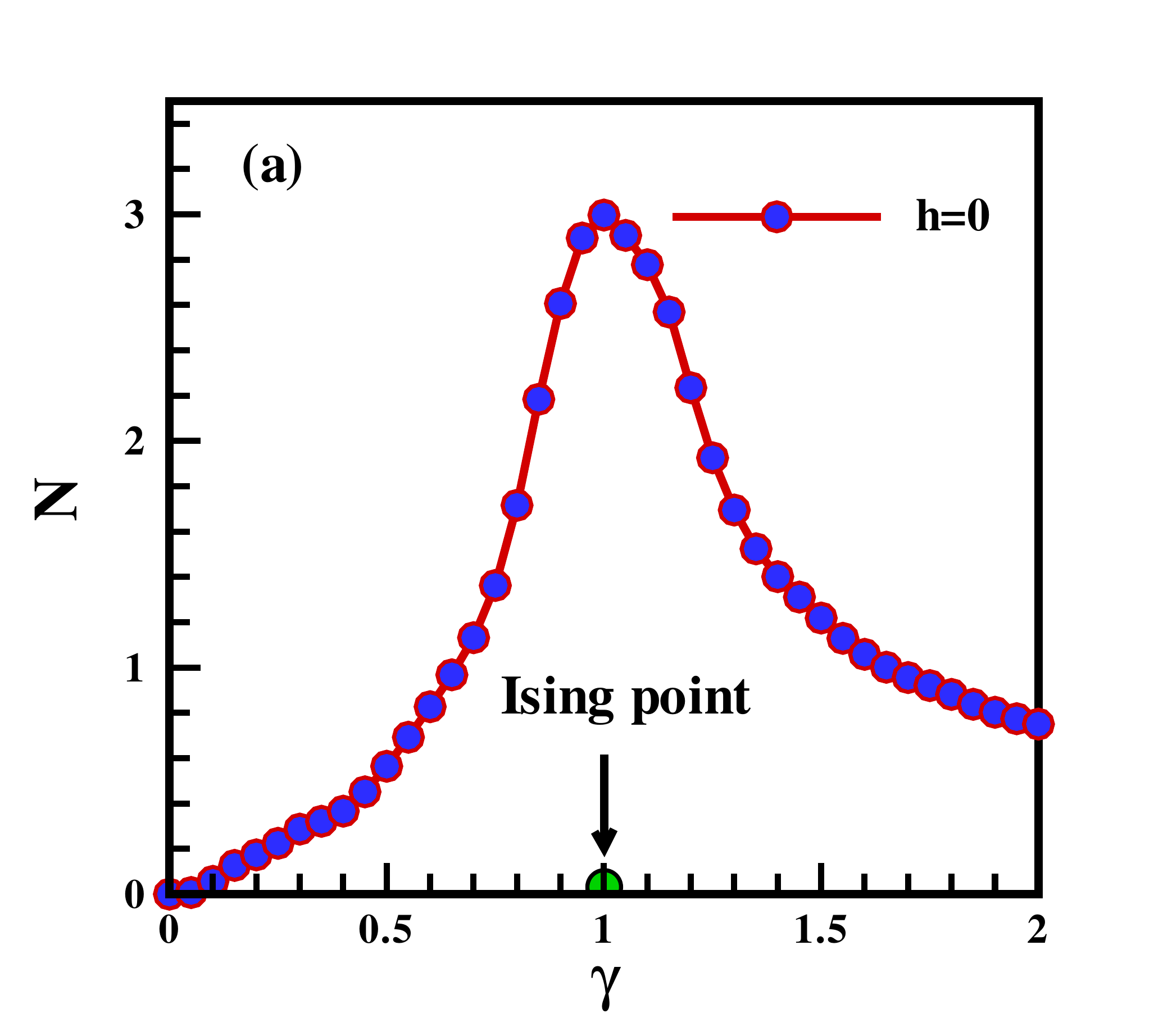,width=1.8in} \psfig{file=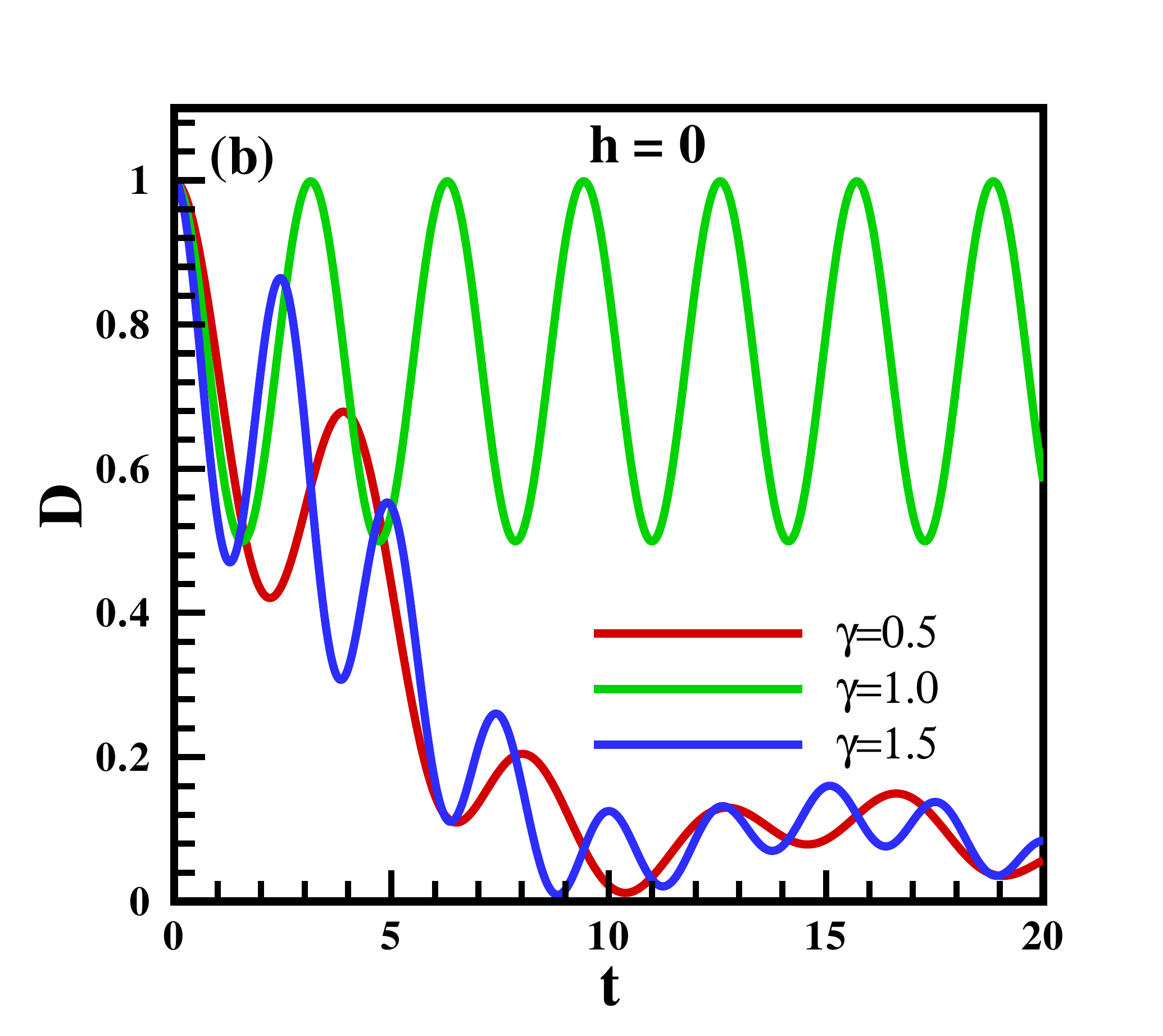,width=1.8in}}
\caption{(color online). (a) The witness of non-Markovianity ${\cal N}$ as a function of the anisotropy parameter $\gamma$. It is clearly seen the non-Markovian dynamics in the region $\gamma>\gamma_c=0$. (b) The trace distance as a function of time for $\gamma=0.5, 1.0, 1.5$. It is completely clear that there are several time intervals where the trace distance increases by time as a trait of the non-Markovian dynamics. It is important to state that the results are obtained for the chain length $N = 6000$ consistent with the thermodynamic limit.}
\label{Fig-1}
\end{figure}
In what follows, we try to determine the dynamical phase diagram of the open quantum system. For this purpose, we consider the initial pair states in which two fermions are created at sites $m$ and $m+1$ in the chain system as 
\begin{eqnarray}
|\psi_1(t=0)\rangle&=&\cos (\phi) a_{m}^{\dagger}|0\rangle+\sin (\phi) a_{m+1}^{\dagger}|0\rangle,\nonumber \\
&=&(\cos (\phi)|10\rangle+\sin (\phi) |01\rangle)_{S} \otimes |0\rangle_{E},\nonumber \\
|\psi_2(t=0)\rangle&=& \cos (\phi') a_{m}^{\dagger}|0\rangle+\sin (\phi') a_{m+1}^{\dagger} |0\rangle,\nonumber \\
&=&(\cos (\phi') |10\rangle+\sin (\phi') |01\rangle)_{S} \otimes |0\rangle_{E}.
\label{initial-state}
\end{eqnarray}
where $|0\rangle$ denotes the vacuum state of the original fermion operators, i.e. $a_k |0\rangle=0$ and $\phi$ is a phase factor.  $|0\rangle_{E}$ is the vacuum state of the environment. One should note that  $\beta_k |0\rangle\neq0$. The relation between the vacuum state of the original fermion operators and the vacuum state of the Bogoliubov operators is given as\cite{Amico11}
\begin{eqnarray}
|0\rangle=\prod_k{(u_k+i v_k \beta^{\dag}_{k}\beta^{\dag}_{-k})}|\Omega\rangle,
\label{vac}
\end{eqnarray}
where $|\Omega\rangle$ is the vacuum state of the Bogoliubov operators. Using Eqs.~(\ref{rel}) and (\ref{vac}), first we calculate the density matrix form of the pair states as a function of time (please see the appendix). Then the matrix form for $\rho_1-\rho_2$ is obtained.  Finally, the trace distance for different values of the phase factors $\phi$ and $\phi'$ is calculated and the optimal state pair is found. In principle, phase factors $\phi$ and $\phi'$ are changed from $0$ up to $2\pi$ and for every set of $\phi$ and $\phi'$ the integral $\int_{\sigma>0}dt~ \sigma(t,\rho_{1,2}(0))$ is calculated and comparing results the optimal state pair is selected. As we have mentioned, the optimal state pair leads to the maximal possible backflow of information during the time evolution of states.
Firstly, we study the effect of the anisotropy parameter $\gamma$ on the dynamical behavior of the system in the absence of the transverse magnetic field, $h=0$. Our results on the witness of non-Markovianity ${\cal N}$ are shown in Fig.~\ref{Fig-1} (a). It is clearly seen that the dynamics of the system is the Markovian in the case of isotropic XX model, $\gamma=0$, in complete agreement with our recent work\cite{Mahmoudi17}. As soon as the anisotropy parameter increases from zero, a dynamical transition happens from the Markovian to the non-Markovian. Thus, in the parameter space of the system, there is a dynamical critical point, $\gamma_c=0$, which this point creates two totally different dynamical behaviors. In addition, the witness of non-Markovianity ${\cal N}$ is maximized exactly at anisotropy parameter  $\gamma=1.0$ where the model is reduced to the Ising model. 
To find a deeper insight into the nature of non-Markovian dynamics where the witness of non-Markovianity ${\cal N}$ is maximized, we concentrated on the time-behavior of the trace distance. Considering initial states as two maximally entangled states, the trace distance for $\gamma=0.5, 1.0, 1.5$  is plotted as a function of time in Fig.~\ref{Fig-1} (b). It is clearly seen that the states are completely distinguishable at $t=0$. In the beginning, when the time increases, the trace distance is decreased for all values of the anisotropy $\gamma$. But by passing the time, the trace distance has different behavior for $\gamma \neq 1.0$ (XY model) and $\gamma=1.0$ (Ising model).
For the values, $\gamma \neq 1.0$, the trace distance will rise and fall irregularly by increasing the time until it gets the value zero at a certain time. After that again the trace distance is increased which is in complete agreement with the non-Markovian dynamics. For Ising point $\gamma=1.0$, the trace distance will fluctuate between $0.5$ and $1.0$, so it will never be zero. Furthermore, surprisingly the initial pair states will be completely distinguishable at some certain times which can be interpreted as  "complete information flow back" to the open quantum system. 
\begin{figure}[t]
\centerline{\psfig{file=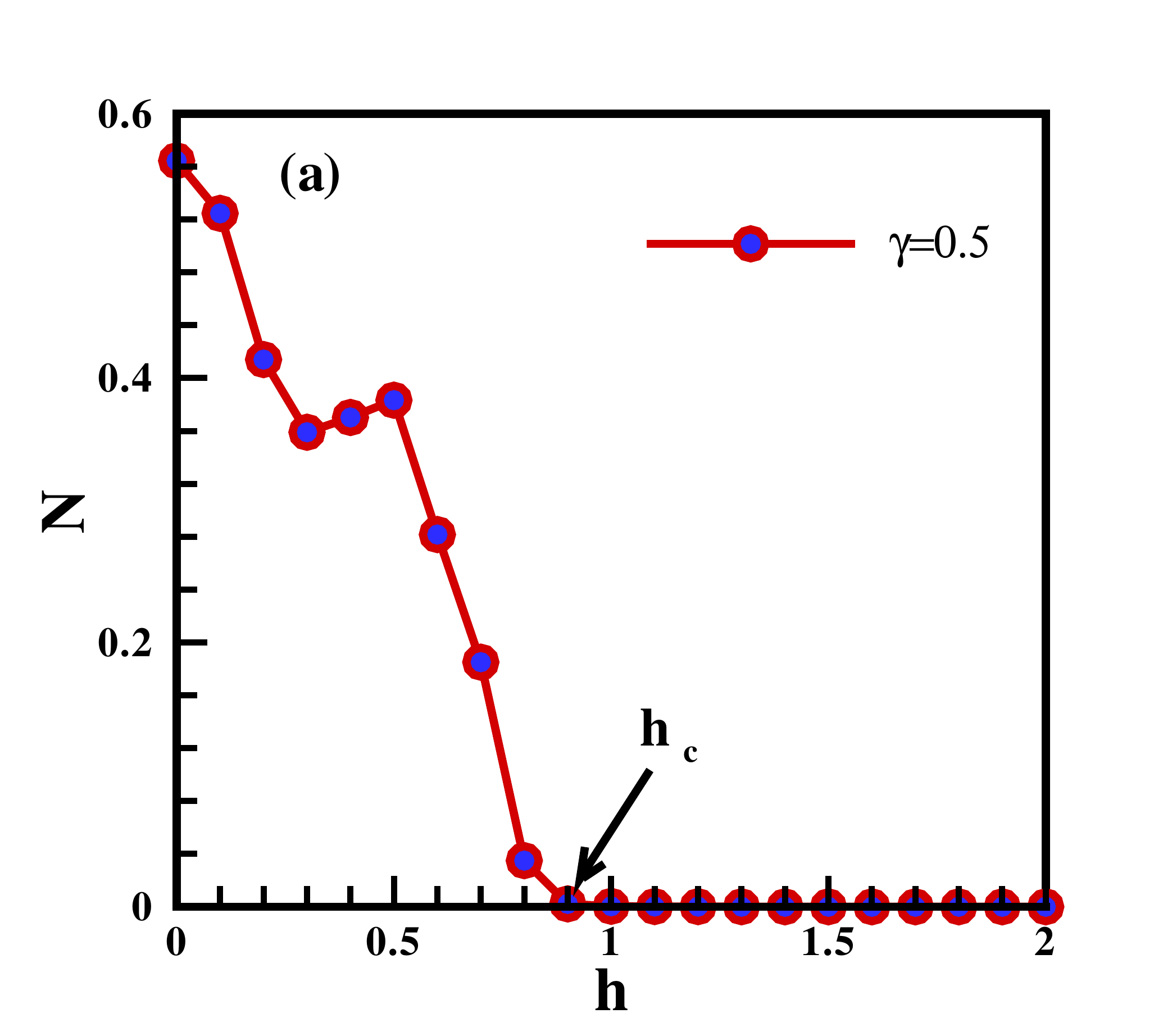,width=1.8in} \psfig{file=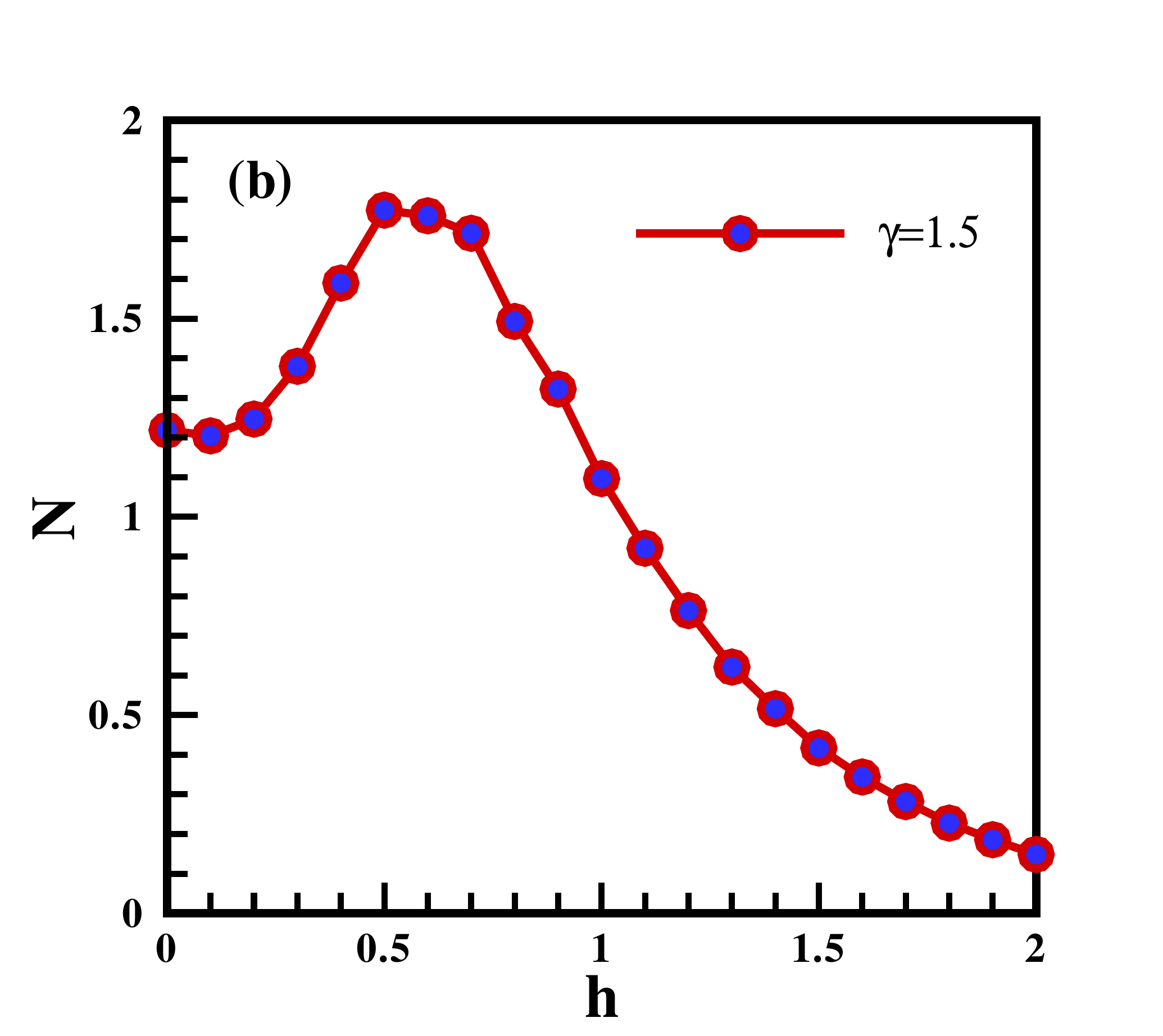,width=1.8in}}
\caption{(color online). The witness of  non-Markovianity ${\cal N}$ as a function of the transverse field $h$ for a value of the anisotropy (a) $\gamma=0.5$ and (b) $\gamma=1.5$. It is clearly seen that a dynamical phase transition into the Markovian dynamics happens at the critical field $h_c \sim 0.9$ when the anisotropy parameter is less than Ising point $\gamma=1.0$. }
\label{Fig-3}
\end{figure}
In the absence of the magnetic field, the system is connected to its environment with the two-point XY interaction. In fact, the two-point Heisenberg interaction between the system and its environment causes flipping spins in the environment, resulting in a flow of the information into the environment. In the XY interaction, $J(1 \pm \gamma)$ are the strength of the flip-flop terms, $S_{n}^{x}S_{n+1}^{x}$ and $S_{n}^{y}S_{n+1}^{y}$. In the isotropic case, $\gamma=0$, the flip-flop term only exchanges the position of neighboring up and down spins as
\begin{eqnarray}
J(S_{n}^{x}S_{n+1}^{x}+S_{n}^{y}S_{n+1}^{y})|\uparrow \downarrow\rangle=\frac{J}{2} |\downarrow \uparrow \rangle.
\label{N_m1}
\end{eqnarray}
Since at $t=0$, all spins of the environment are aligned in the opposite of the $Z$ axis,  in the other words they are paralleled, the isotropic XY interaction has no effect on the spins of the environment except those on the edges. On the other hand, as soon as the anisotropy is applied, $\gamma \neq 0$, the XY interaction changes the direction of neighboring parallel spins  in addition to exchanging the position of neighboring up and down spins as
\begin{eqnarray}
(J(1+\gamma)S_{n}^{x}S_{n+1}^{x}+J(1-\gamma) S_{n}^{y}S_{n+1}^{y})|\uparrow \uparrow\rangle&=&\frac{J\gamma}{2} |\downarrow \downarrow \rangle \nonumber\\
(J(1+\gamma)S_{n}^{x}S_{n+1}^{x}+J(1-\gamma) S_{n}^{y}S_{n+1}^{y})|\downarrow \downarrow\rangle&=&-\frac{J\gamma}{2} |\uparrow \uparrow \rangle, \nonumber\\
\label{N_m2}
\end{eqnarray}
which explicitly shows that the anisotropic XY interaction has a significant effect on all spins of the environment, resulting in a flow back of the information into the system.
Secondly, we investigate the effect of the transverse magnetic field on the dynamics of the system. As we have mentioned, the dynamics of a qubit interacting with the same environment has been studied\cite{Appolaro11}. It has been shown that a critical field exists in the parameter space, where the qubit dynamics is effectively Markovian and such critical magnetic field separates two regions with completely different dynamical behaviors. Here we consider the nearest neighbor pair spins as an open quantum system. The results of such an analysis are shown in Fig.~\ref{Fig-3}. The witness of non-Markovianity ${\cal N}$ is plotted as a function of the transverse field $h$ for different anisotropy parameter (a) $\gamma=0.5$ and (b) $\gamma=1.5$. As it is seen in Fig.~\ref{Fig-3} (a), the witness of non-Markovianity ${\cal N}$ is decreased almost monotonically by increasing the transverse field and will be exactly zero at $h_c(\gamma=0.5)=0.9$, which shows that the transverse field can create a dynamical transition from the non-Markovian to the Markovian. In the region $\gamma>1.0$, Fig.~\ref{Fig-3} (b), at first the witness of non-Markovianity ${\cal N}$ is increased by applying the transverse field. It will be maximized at a certain value of the transverse field. More increasing the transverse field, ${\cal N}$ is decreased monotonically and behaves asymptotically with respect to the time.

\section{Loschmidt echos}\label{sec4}
\begin{figure}[t]
\centerline{\psfig{file=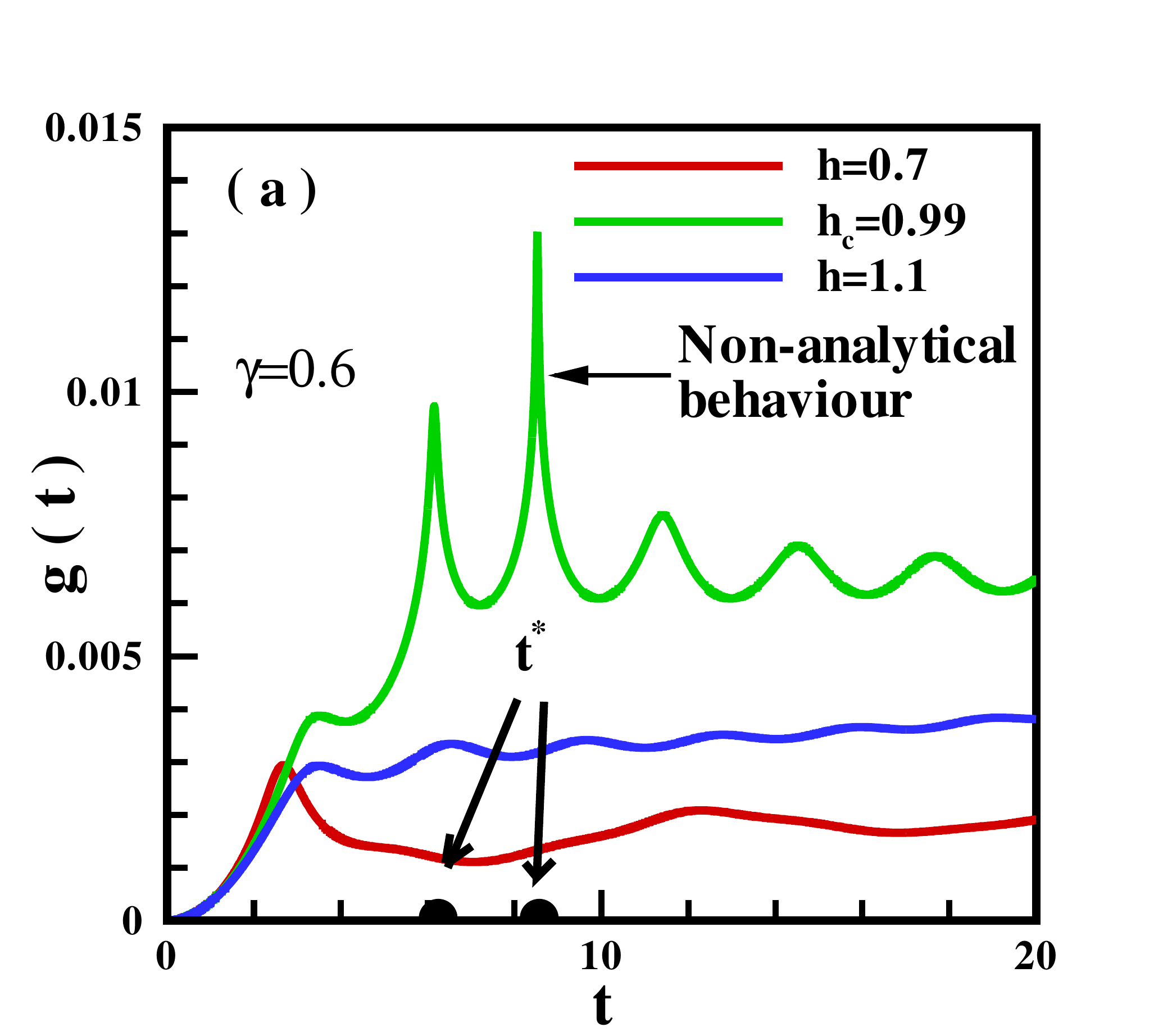,width=1.8in} \psfig{file=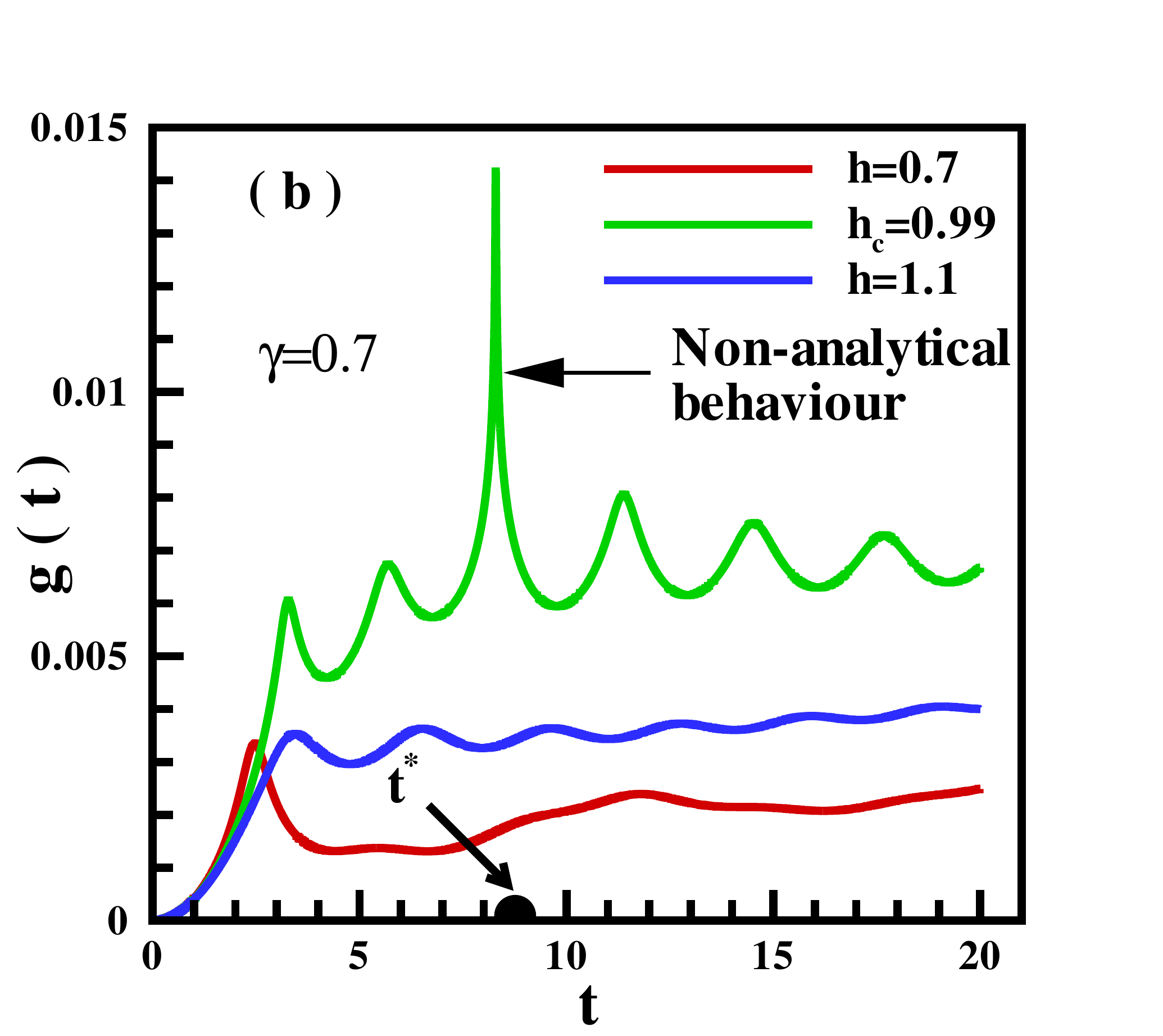,width=1.8in}}
\centerline{\psfig{file=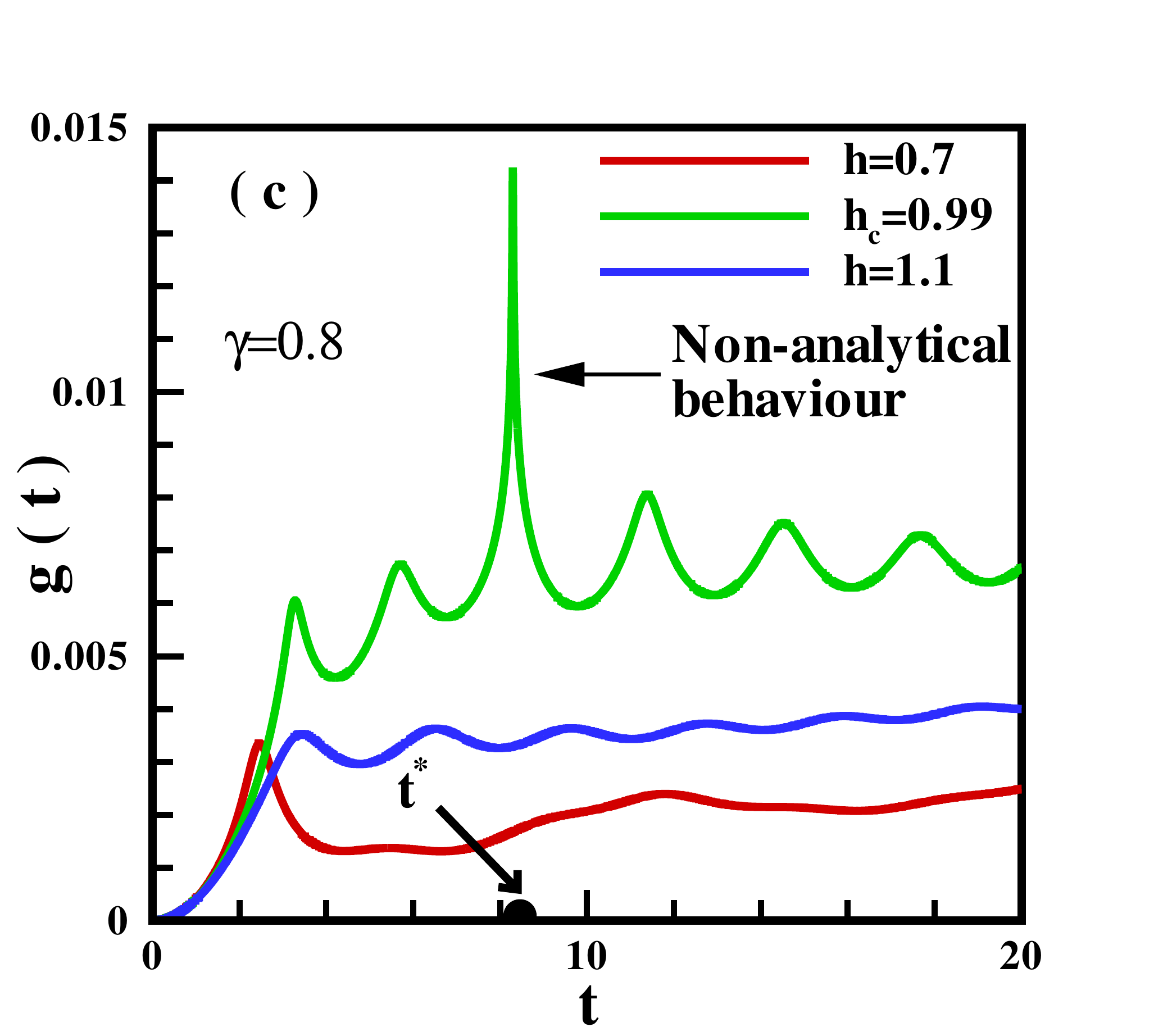,width=1.8in}}
\caption{(color online).  The rate function $g(t)$ as a function of $t$ (for a short time) from $\gamma=0.6$ to $\gamma=0.8$ for three values of applied magnetic field. For each figure, the red, green, and blue lines represent $h<h_c$, $h=h_c$ and $h>h_c$, respectively.}
\label{Fig-4}
\end{figure}
In this section, a closed quantum many-body system has been investigated. In fact, the system and its environment in the previous section are considered as a closed quantum system. We are interested to observe the amplitude of overlap of the time-evolved state with the initial state. This is also called the Loschmidt amplitude. Recently it has been shown that the nonequilibrium real-time evolution after a quantum quench can generate non analyticities as a function of time in the time-evolved state with initial state\cite{Heyl13, Heyl17}. Here, we show that in the Markovian to the non-Markovian dynamical phase transition, a non-analytic behavior in time can be observed in the time-evolved of the Loschmidt echo.\\
The Loschmidt amplitude is presented as 
\begin{eqnarray}
G(t)=\langle \psi_0 \vert \psi(t) \rangle =\langle \psi_0 \vert e^{-i H t}\vert \psi_0  \rangle. 
\label{fidelity}
\end{eqnarray}
The related probability is known as Loschmidt echo,
\begin{eqnarray}
L(t)=\vert G(t)\vert^2.
\label{LE}
\end{eqnarray}
Using these two items, various cases can be examined such as return amplitudes. Loschmidt amplitudes are similar to the partition functions and both of them depend on the number of degrees of freedom $N$
\begin{eqnarray}
G(t)=e^{-Ng(t)},
\label{exp(g(t))}
\end{eqnarray}
where $g(t)$ is the Loschmidt-echo return rate function. The above equation can be rewritten
\begin{eqnarray}
g(t)=-lim _{N\rightarrow\infty}\frac{1}{N}log(G(t)).
\label{g(t)}
\end{eqnarray}
As we know, the observation of a non-analytic behavior at the phase transition point is expected. In fact, we are looking for the time that the physical state of the system has the least similarity to the physical state of the system at $t=0$. In this case, we see non-analytic behavior at certain times in $g(t)$ which are called critical times. 
We have calculated $g(t)$ for different values of the anisotropy $\gamma=0.6, 0.7, 0.8$. Results are presented in Figs.~\ref{Fig-4}. As it is shown in Figs.~\ref{Fig-4}, the Loschmidt-echo return rate exhibits non-analyticity in evolution time only at the critical field $h_c$, where the Markovian to the non-Markovian dynamical transition occurs. It is clearly seen that the critical time, $t^{\star}$, where the non-analytic behavior in the Loschmidt-echo return rate function happens, depends on the anisotropy parameter $\gamma$.

\section{Conclusion}\label{sec5}
In this paper, we have considered the anisotropic spin-1/2 XY chain in the presence of a transverse magnetic field. By selecting a two-spin system from the spin chain as an open quantum system, the rest of the chain plays the role of an environment. The desired open quantum system can be coupled to the environment via two-spin XY Heisenberg interaction. For recognizing the degree of non-Markovian behavior in the open quantum system, we used a powerful measure which is based on the trace distance. 
We showed that in the absence of the transverse magnetic field, the dynamics of the open quantum system is Markovian when the two-point Heisenberg interaction is isotropic. As soon as the anisotropy is applied, a dynamical transition into the non-Markovian region occurs. We argued that, since the two-point anisotropic Heisenberg interaction can induce the excitation in the whole range of the environment, a flow back information into the system is possible which is known as the non-Markovian behavior. In addition, all the information flows back from the environment to the system at the certain Ising value of the anisotropy $\gamma=1.0$.
In the presence of the transverse magnetic field, the situation is different. Since the Zeeman term tries to align all spins of the chain, it is natural to expect a dynamical transition into the Markovian region. We showed that the field induced the Markovian into the non-Markovian transition only in the region of anisotropy parameter $\gamma \leq 1.0$. In fact, in the region of  $\gamma > 1.0$, the quantum fluctuations are strong enough and do not allow the formation of a quantum state with all paralleled spins.  
Finally, we focused on the Loschmidt echo. We calculated the Loschmidt-echo return rate function in the thermodynamic limit of our chain model.  Results showed that the dynamical transition from Markovian into the non-Markovian is revealed as a non-analyticity in the real-time evolution of the Loschmidt-echo return rate function.

\section{ACKNOWLEDGMENT}
The authors wish to thank R. Jafari for useful comments and discussions. 
\section{Appendix}
Here we try to calculate $X^{+}_{mm'}$ which is related to the fermion operators as 
\begin{eqnarray}
X^{+}_{mm'}&=&\langle a^{\dag}_{m}(t) a_{m}(t)\rangle \langle a^{\dag}_{m'}(t) a_{m'}(t)\rangle   \nonumber\\
&+&\langle a^{\dag}_{m}(t) a_{m'}(t)\rangle \langle a_{m}(t)a^{\dag}_{m'}(t)\rangle. 
\end{eqnarray}
At first, the method of calculating $\langle a^{\dag}_{m}(t) a_{m}(t)\rangle$ is explained. 
\begin{eqnarray}
\langle a^{\dag}_{m}(t) a_{m}(t)\rangle &=& \langle\psi_0|e^{iHt}a^{\dag}_{m}a_{m}e^{-iHt}|\psi_0\rangle \nonumber \\
&=&\frac{1}{\sqrt{N}}\sum _{k}e^{i(k_1-k_2)m}\nonumber \\
&\langle&\psi_0|e^{iHt}a^{\dag}_{k_1}a_{k_2}e^{-iHt}|\psi_0\rangle 
\label{N_m}
\end{eqnarray}
\begin{eqnarray}
e^{iHt}a^{\dag}_{k_1}e^{-iHt} &=& e^{it\sum_{k}\varepsilon(k) (\beta_{k}^{\dagger}\beta_{k})/\hbar}(cos(k_1)\beta_{k_1}^{\dagger} \nonumber\\
&+&i sin(k_1)\beta_{-k_1})e^{-it\sum_{k}\varepsilon(k) (\beta_{k}^{\dagger}\beta_{k})/\hbar} \nonumber\\
&=& cos(k_1)e^{it\varepsilon(k_1)}\beta_{k_1}^{\dagger}\nonumber \\
&+&i sin(k_1)e^{-it\varepsilon(-k_1)}\beta_{-k_1} 
\end{eqnarray}
In the same way one can show
\begin{eqnarray}
e^{iHt}a_{k_2}e^{-iHt} &=& cos(k_2)e^{-it\varepsilon(k_2)}\beta_{k_2}^{\dagger}\nonumber \\
&-&isin(k_2)e^{it\varepsilon(-k_2)}\beta_{-k_2}^{\dagger}.
\end{eqnarray}
The relation \ref{N_m} is simplified as follows
\begin{eqnarray}
&\langle& a^{\dag}_{m}(t) a_{m}(t)\rangle =\frac{1}{\sqrt{N}}\sum _{k}e^{i(k_1-k_2)m}\nonumber\\
&\langle&\psi_0|(cos(k_1)e^{it\varepsilon(k_1)}\beta_{k_1}^{\dagger}\nonumber \\
&+&i sin(k_1)e^{-it\varepsilon(-k_1)}\beta_{-k_1}) \nonumber\\
&(&cos(k_2)e^{-it\varepsilon(k_2)}\beta_{k_2}^{\dagger}\nonumber \\
&-&isin(k_2)e^{it\varepsilon(-k_2)}\beta_{-k_2}^{\dagger})|\psi_0\rangle. \nonumber \\
\label{N1_m}
\end{eqnarray}
Using the Bogoliubov operators
\begin{eqnarray}
\langle a^{\dag}_{m}(t) a_{m}(t)\rangle &=& \frac{1}{2 N^2}\sum _{k,k'}(1+e^{i(k+\phi)}+e^{-i(k'+\phi))}\nonumber \\
&+&e^{i(k-k')})\nonumber\\
&(&e^{-it(\varepsilon(k)-\varepsilon(k'))}cos^2(k)cos^2(k')\nonumber \\
&+&e^{it(\varepsilon(k)+\varepsilon(k'))}sin^2(k)cos^2(k')\nonumber\\
&+&e^{-it(\varepsilon(k)+\varepsilon(k'))}sin^2(k')cos^2(k)\nonumber\\
&+& e^{it(\varepsilon(k)-\varepsilon(k'))}sin^2(k)sin^2(k')\nonumber\\
&+&\frac{1}{4}sin(2k)sin(2k')(-e^{it(\varepsilon(k)-\varepsilon(k'))}\nonumber\\
&+&e^{it(\varepsilon(k)+\varepsilon(k'))}\nonumber\\
&+&e^{-it(\varepsilon(k)+\varepsilon(k'))}-e^{-it(\varepsilon(k)-\varepsilon(k'))}))\nonumber\\
&+& \frac{1}{4 N^2}\sum _{k,k'}(sin^2(2k') (1+cos(k+\phi))\nonumber\\
&(&2-e^{2it(\varepsilon(k')}-e^{-2it(\varepsilon(k')}) \nonumber
\end{eqnarray}
Similarly, $\langle a^{\dag}_{m}(t) a_{m+1}(t)\rangle$ can be written as follows
\begin{eqnarray}
\langle a^{\dag}_{m}(t) a_{m+1}(t)\rangle &=&\frac{1}{2 N^2}\sum _{k,k'}(1+e^{i(k+\phi)}+e^{-i(k'+\phi))}\nonumber \\
&+&e^{i(k-k')}) e^{-ik}\nonumber\\
&(&e^{-it(\varepsilon(k)-\varepsilon(k'))}cos^2(k)cos^2(k')\nonumber \\
&+&e^{it(\varepsilon(k)+\varepsilon(k'))}sin^2(k)cos^2(k')\nonumber\\
&+&e^{-it(\varepsilon(k)+\varepsilon(k'))}sin^2(k')cos^2(k)\nonumber \\
&+&e^{it(\varepsilon(k)-\varepsilon(k'))}sin^2(k)sin^2(k')\nonumber\\
&+&\frac{1}{4}e^{ik'}sin(2k)sin(2k')(-e^{it(\varepsilon(k)-\varepsilon(k'))}\nonumber \\
&+&e^{it(\varepsilon(k)+\varepsilon(k'))}\nonumber\\
&+&e^{-it(\varepsilon(k)+\varepsilon(k'))}-e^{-it(\varepsilon(k)-\varepsilon(k'))}))\nonumber \\
&+& \frac{1}{4 N^2}\sum _{k,k'}(sin^2(2k')\nonumber\\
&(&1+cos(k+\phi))e^{-ik'}(2-e^{2it(\varepsilon(k')}-e^{-2it(\varepsilon(k')}) \nonumber \\
\end{eqnarray}
\\
\begin{figure}[t]
\centerline{\psfig{file=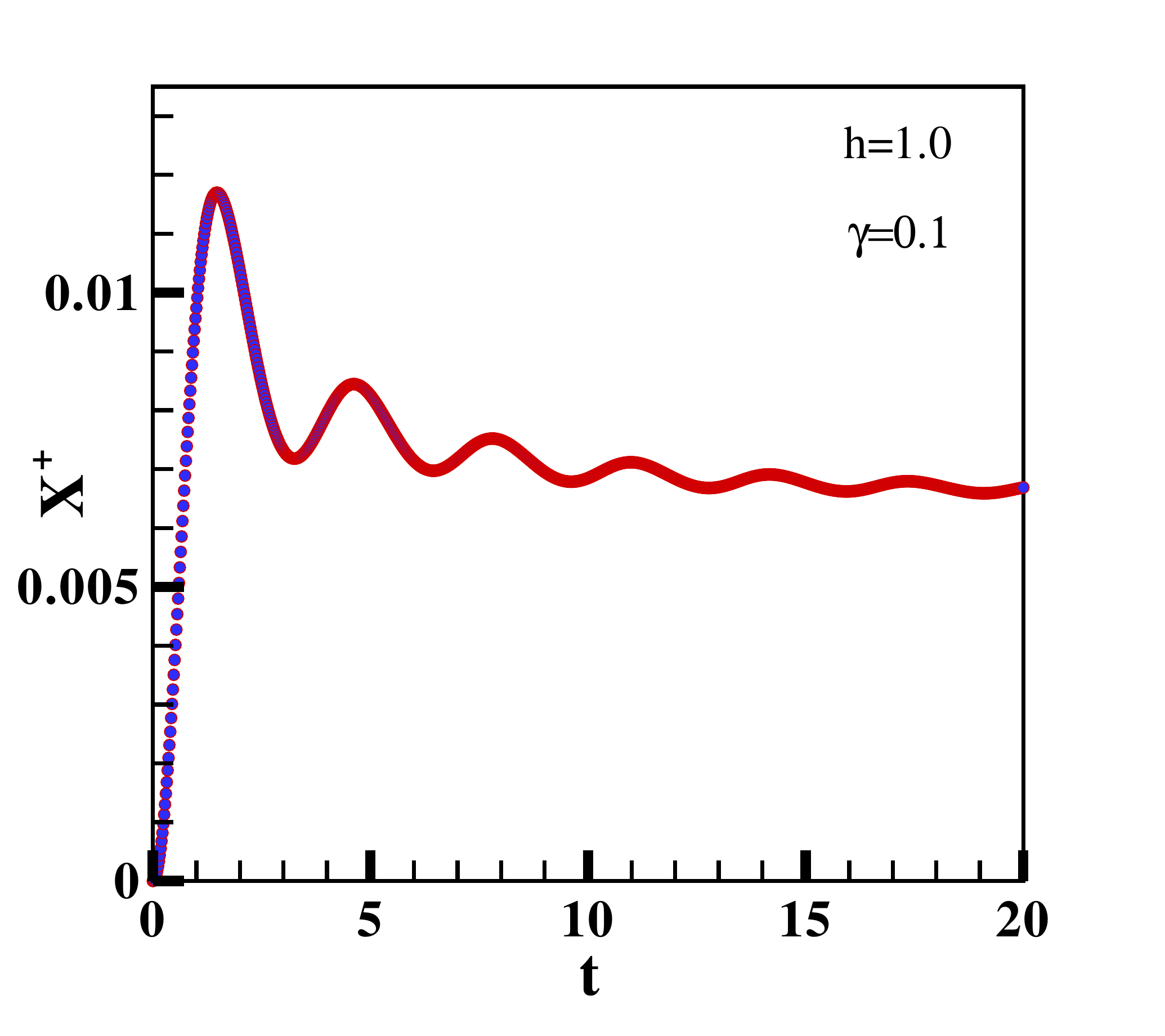,width=3in}}
\caption{(color online). The function $X^{+}$ in terms of time.}
\label{Fig-5}
\end{figure}
\\
\vspace{0.3cm}


\end{document}